\documentstyle[12pt]{article}
\begin{document}

\title{Generally relativistical Tetrode-Weyl-Fock-Ivanenko formalism
       and behaviour of quantum-mechanical particles of spin $1/2$
       in the Abelian monopole field.}

\author {V.M.Red'kov \\
     Institute of Physics, Belarus Academy of Sciences\\
F Skoryna  Avenue 68, Minsk 72, Republic ob Belarus\\
e-mail: redkov@dragon.bas-net.by}

\maketitle


\begin{abstract}

Some attention in the literature has been given to the case
of a~particle of spin $1/2$ on the background of   the~external
monopole potential. Certain aspects of this problem are reexamined here.
The~primary technical `novelty' is that the~tetrad generally
relativistic method of Tetrode-Weyl-Fock-Ivanenko for
describing a~spinor particle is exploited.
The choice of the~formalism to deal with the~monopole-doublet  problem
has turned out to be of great  fruitfulness for examining this system.
It is matter that,
as known, the~use of a~special spherical tetrad
in the~theory of a~spin $1/2$ particle had led Schr\"{o}dinger
to a~basis of remarkable features. In particular,
the~following explicit expression for momentum operator components
had been calculated
$ J_{1} = l_{1} + i \sigma ^{12} \cos \phi / \sin\theta
\; , \;
J_{2}= l_{2} + i \sigma ^{12} \sin \phi / \sin\theta
\; , \; J_{3} = l_{3}$ .
This basis has been used with great
efficiency by Pauli in his  investigation on  the~problem of allowed
spherically symmetric wave  functions in quantum mechanics.  For our purposes,
just several simple  rules extracted from the~much more comprehensive
Pauli's analysis will be quite   sufficient; those are almost mnemonic
working regulations. So, one may
remember some very primary facts of $D$-functions theory and then produce,
almost automatically, proper wave functions. It  seems rather likely,
that there may exist a~generalized analog of such a~representation
for $J_{i}$-operators, that  might be successfully used whenever
in a~linear problem there exists a~spherical symmetry, irrespective of
the~concrete embodiment of such a~symmetry.  In particular, the~case of
electron in the~external Abelian monopole field, together with
the~ problem of selecting the~allowed wave functions as well as
the~Dirac charge quantization condition,  completely come under
that Shr\"{o}dinger-Pauli method.
In particular,  components of the generalized
conserved momentum can be expressed as follows
$
j^{eg}_{1} = l_{1} + (i\sigma ^{12} - eg) \cos \phi / \sin \theta  \; ,
\; j^{eg}_{2} = l_{2} + (i\sigma ^{12} - eg) \sin \phi / \sin \theta \; , \;
j^{eg}_{3} = l_{3}$,
where $e$ and $g$  are an~electrical and magnetic charge,
respectively. In accordance with the~above regulations,
the~corresponding electron-monopole wave functions can be
constructed like in the purely electron pattern but witn a
single change
$
D^{j}_{-m,\pm1/2}(\phi ,\theta ,0) \;\; \rightarrow \;\;
D^{j}_{-m,eg\pm1/2}(\phi ,\theta ,0) \; $.
\end{abstract}

\newpage
\subsection*{1. Introduction.}

While  there  not  exists  at  present  definitive  succeeded
experiments concerning  monopoles, it is nevertheless true  that
there exists a~veritable jungle  of  literature  on  the~monopole
theories. Moreover, properties of more  general
monopoles, associated with large gauge groups now thought to be
relevant in  physics.
As evidenced  even  by   a~cursory examination of some popular surveys
(see, for example, [1,2]),  the~whole monopole area covers and touches
quite a~variety of fundamental problems. The~most outstanding of them are:
the~electric charge quantization [3-10], $P$-violation in purely  electromagnetic
processes [11-16], scattering on the~Dirac string [17-19], spin from monopole
and spin from isospin [20-23], bound states in fermion-monopole system
and violation of the Hermiticity property [24-38],
fermion-number breaking in the~presence of a~magnetic monopole and
monopole catalysis of baryon decay [39-41].

The~tremendous volume of publications on monopole topics
(and there is no hint  that  its  raise  will  stop)  attests
the~interest which  they  enjoy among theoretical  physicists,  but
the~same token, clearly  indicates  the~unsettled  and  problematical
nature of those objects:  the~puzzle of monopole seems to be
one of  the~still yet unsolved problems of particle  physics\footnote{Very
physicists have contributed to investigation of the~monopole-based theories.
The~wide scope of the~field and the~prodigious number of investigators
associated with various of its developments make it all but hopleless to list
even the~principal contributors.
The present study does not pretend to be
a~survey in this matter, so I give but a few of the~most important references
which may be useful to the~readers who wish some supplementary material or
are interested in more techical developments beyonds the~scope of the~present
treatment.}
In general, there are several ways of approaching  the monopole problems.
As known, together with geometrically topological  way  of  exploration into
them,  another  approach   to   studying  such
configurations is possible; namely, that  concerns  any physical
manifestations  of monopoles when they are considered as external  potentials.
Moreover,
from the~physical standpoint, this latter method can be thought  of  as
a~more visualizable one in comparison with  less  obvious  and  more
direct  topological  language.

Some more concrete  remarks  referring  to  our  further
work and designated to delineate its content are to be given.
The most attention in the literature has been given to the case
of a~particle of spin $1/2$ on the background of   the~external
monopole potential: for the~Abelian  case see, for instance, [42-45];
for the non-Abelian one see [46-54].
For the present work,  the Abelian situation  only will be treated;
on the line developed here a corresponding non Abelian system   will
be considered in a separate work.

Now, for convenience of the readers, some remarks about the~approach and
technique used in the~work are to be given.
The~primary technical `novelty' is that, in the paper, the~tetrad (generally
relativistic) method [55-63] of Tetrode-Weyl-Fock-Ivanenko (TWFI) for
describing a~spinor particle will be exploited.
 The choice of the~formalism to deal with the~monopole-doublet  problem
has turned out to be of great  fruitfulness for examining this system.
Taking of just this method is  not an~accidental step. It is matter that,
as known (but seemingly not very vastly), the~use of a~special spherical tetrad
in the~theory of a~spin $1/2$ particle had led Schr\"{o}dinger and
 Pauli [64, 65]  to a~basis of remarkable features. In particular,
the~following explicit expression for (spin $1/2$ particle's)
momentum operator components had been calculated
$$
J_{1}= l_{1} + {{i \sigma ^{12} \cos \phi }\over{ \sin\theta}}
\; , \qquad
J_{2}= l_{2} + {{i \sigma ^{12} \sin \phi }\over{\sin\theta}}
\; , \qquad J_{3} = l_{3}
\eqno(1.1)
$$

\noindent   just that kind of structure for $J_{i}$ typifies this frame
in bispinor space. This Schr\"{o}dinger's basis had been used with great
efficiency by Pauli
in his  investigation [65] on   the~problem of allowed spherically
symmetric wave  functions in quantum mechanics.  For our purposes,
just several simple
rules extracted from the~much more comprehensive Pauli's analysis will be
quite   sufficient (those are almost mnemonic working regulations).
They can be explained on the~base of $S=1/2$ particle case.
To this end, using any representation of $\gamma$ matrices where
$\sigma^{12} = {1 \over 2}\;(\sigma_{3} \oplus \sigma_{3})$  (throughout the work,
the Weyl's spinor  frame is used) and taking into account the~explicit form for
$\vec{J}^{2}, J_{3}$ according to (1.1), it is readily verified that the~most
general bispinor functions  with fixed quantum numbers $j,m$ are to be
 (see also in [61])
$$
\Phi _{jm}(t,r,\theta ,\phi ) =
                              \left ( \begin{array}{l}
          f_{1}(t,r) \; D^{j}_{-m,-1/2}(\phi ,\theta ,0) \\
          f_{2}(t,r) \; D^{j}_{-m,+1/2}(\phi ,\theta ,0) \\
          f_{3}(t,r) \; D^{j}_{-m,-1/2}(\phi ,\theta ,0) \\
          f_{4}(t,r) \; D^{j}_{-m,+1/2}(\phi ,\theta ,0)
                               \end{array} \right )
\eqno(1.2)
$$

\noindent where $D^{j}_{mm'}$ designates the Wigner's $D$-functions
(the~notation and subsequently required formulas according to [66], are
adopted).
One should take notice of the low right indices
$-1/2$ and $+1/2$ of $D$-functions  in (1.2),  which correlate with the~explicit diagonal structure of
the~matrix $\sigma^{12} = {1 \over 2}\; ( \sigma_{3} \oplus \sigma_{3})$.
The~Pauli criterion allows only half integer values for $j$.

So, one may
remember some very primary facts of $D$-functions theory and then produce,
almost automatically, proper wave functions. It  seems rather likely,
that there may exist a~generalized analog of such a~representation
for $J_{i}$-operators, that  might be successfully used whenever
in a~linear problem there exists a~spherical symmetry, irrespective of
the~concrete embodiment of such a~symmetry.  In particular, the~case of
electron in the~external Abelian monopole field, together with
the~ problem of selecting the~allowed wave functions as well as
the~Dirac charge quantization condition,  completely come under
that Shr\"{o}dinger-Pauli method. In particular,  components of the generalized
conserved momentum can be expressed as follows (for more detail, see [67])
$$
j^{eg}_{1} = l_{1} + {{(i\sigma ^{12} - eg) \cos \phi}
\over { \sin \theta }} \; ,  \qquad
j^{eg}_{2} = l_{2} + {{(i\sigma ^{12} - eg) \sin\phi}
\over {\sin \theta }} \; ,  \qquad
j^{eg}_{3} = l_{3}
\eqno(1.3)
$$

\noindent where $e$ and $g$  are an~electrical and magnetic charge,
respectively. In accordance with the~above regulations,
the~corresponding electron-monopole wave functions can be
constructed like in the purely electron pattern (1.2) but witn a
single change
$$
D^{j}_{-m,\pm1/2}(\phi ,\theta ,0) \;\; \rightarrow \;\;
D^{j}_{-m,eg\pm1/2}(\phi ,\theta ,0) \; .
\eqno(1.4)
$$

\noindent The~Pauli criterion produces two results: first,
$\mid eg \mid = 0, 1/2, 1, 3/2,\ldots $ (what is called the~Dirac charge
quantization condition; second, the quantum number $j$
in (1.4) may take the values $\mid eg \mid -1/2 , \mid eg \mid +1/2,
\mid eg \mid +3/2, \ldots$ that  selects the~proper spinor particle-monopole
functions.

There exists additional line justified the~interest to just
the~aforementioned approach: the~Shr\"{o}dinger's tetrad basis and
Wigner's $D$-functions are deeply  connected with what is called
 the~formalism of spin-weight harmonics [68-70]  developed in
the~frame of the~Newman-Penrose method of light (or isotropic))
tetrad. Some relationships between spin-weight and spinor monopole
harmonics have already been examined in the literature [71-73],
the~present work   follows the~notation used in [67].

\subsection*{2. The Pauli criterion.}
Let the $J^{\lambda }_{i}$   denote
$$
J_{1}= (\; l_{1} + \lambda\; {{\cos  \phi  }\over {\sin \theta}}\; ),\qquad
J_{2}= (\; l_{2} + \lambda\; {{\sin  \phi  }\over {\sin \theta}}\; ),\qquad J_{3} = l_{3}
\eqno(2.1)
$$

\noindent at an~arbitrary $\lambda$,  as readily  verified, those $J_{i}$
 satisfy the~commutation rules of the~Lie algebra
$SU(2): [ J_{a},\; J_{b} ] = i \; \epsilon _{abc} \; J_{c}$.
As known,  all  irreducible  representations  of  such  an~ abstract
algebra are determined by a~set of weights
   $j = 0, 1/2, 1, 3/2,... \; \; ({\em dim} \;j = 2j + 1)$.
Given the~explicit expressions of  $J_{a}$   above,  we  will
find functions
               $\Phi ^{\lambda }_{jm}(\theta ,\phi )$
on which the~representation of  weight $j$ is realized. In agreement with
the~general approach [65], those solutions are to be established by
the~following relations
$$
J_{+} \; \Phi ^{\lambda }_{jj} \; = \; 0 ,\qquad
\Phi ^{\lambda }_{jm} \; = \; \sqrt{{(j+m)! \over (j-m)! \; (2j)! }} \; J^{(j-m)}_{-} \;
\Phi^{\lambda}_{jj}  ,
\eqno(2.2)
$$
$$
J_ {\pm} \; = \; ( J_{1} \pm i J_{2}) \; = \;
e^{\pm i\phi }\; [\; \pm { \partial \over \partial \theta } \; + \;
 i \cot \theta \; { \partial \over  \partial \phi} \; + \;
 { \lambda \over  \sin  \theta }\; ] .
$$

\noindent From the equations
$J_{+} \; \Phi ^{\lambda }_{jj} \; = \; 0 \;$ and
$\; J_{3} \; \Phi ^{\lambda }_{jj} \;  = \; j \; \Phi ^{\lambda }_{jj}$
it follows that
$$
\Phi ^{\lambda }_{jj}  = \;
N^{\lambda }_{jj} \;  e^{ij\phi} \; \sin^{j}\theta \;\;
{( 1 + \cos \theta  )^{+\lambda /2} \over ( 1 - \cos \theta )^{\lambda /2}} ,\;
N^{\lambda }_{jj} \; =  \; {1 \over \sqrt{2\pi}} \; { 1 \over 2^{j} } \;
 \sqrt{{(2j+1) \over \Gamma(j+m+1) \; \Gamma(j-m+1)}} .
$$

\noindent Further, employing (2.2) we produce the~functions
$\Phi ^{\lambda }_{jm}$
$$
\Phi ^{\lambda }_{jm} \; = \; N^{\lambda }_{jm} \;  e^{im\phi} \;
 {1 \over \sin^{m}\theta }
{(1 - \cos \theta)^{\lambda/2} \over (1 + \cos \theta)^{+\lambda/2}} \; \times
$$
$$
({ d \over d \cos  \theta})^{j-m} \; [\; (1 + \cos  \theta ) ^{j + \lambda } \;
(1 - \cos  \theta ) ^{j-\lambda } \; ]
\eqno(2.3)
$$

\noindent where
$$
N^{\lambda }_{jm} \;  = \; {1 \over \sqrt{2\pi} 2^{j}} \;
 \sqrt{{(2j+1) \; (j+m)! \over
2(j-m)!  \Gamma(j + \lambda +1) \; \Gamma(j- \lambda +1)}}
$$

\noindent The Pauli criterion tells us that the $(2j + 1)$ functions
$ \Phi ^{\lambda }_{jm}(\theta ,\phi ), \; m = - j,... , +j$
so  constructed are  guaranteed  to be a~basis for a~finite-dimension
representation, providing that the function
$\Phi ^{\lambda }_{j,-j}(\theta ,\phi )$
found by this procedure obeys the~identity
$$
J_{-} \;\; \Phi ^{\lambda }_{j,-j} \; = \; 0 \; .
\eqno(2.4a)
$$

\noindent After substituting the~function
 $\Phi ^{\lambda }_{j,-j}(\theta ,\phi )$
(in  the~form  given  (2 3))  to the (2.4a), the latter reads
$$
J_{-} \; \Phi ^{\lambda }_{j,-j} \; = \;  N^{\lambda }_{j,-j} \;
e^{-i(j+1)\phi }\; (\sin \theta)^{j+1}  \;
{(1 - \cos  \theta )^{\lambda /2} \over (1 + \cos  \theta )^{\lambda /2}} \; \times
$$
$$
({d \over d \cos \theta})^{2j+1} \; [\; (1 + \cos  \theta )^{j+\lambda } \;
(1 - \cos  \theta )^{j-\lambda } ) \; ]\;  = \;  0
\eqno(2.4b)
$$

\noindent which in turn gives the~following restriction on
 $j$  and $\lambda $
$$
({d \over d \cos  \theta})^{2j+1} \; [\; (1 + \cos  \theta  )^{j+\lambda } \;
(1 - \cos  \theta  )^{j-\lambda } \; ] \;  = \; 0.
\eqno(2.4c)
$$

\noindent But the~relation (2.4c) can be satisfied  only  if  the~factor
 $P(\theta )$
subjected to the~operation of taking derivative
$( d/d \cos \theta ) ^{2j+1}$
is a~polynomial of degree $2j$  in $ \cos \theta$. So, we have (as a~result
of the~Pauli criterion)

1.  {\em the} $\lambda$ {\em is allowed to take values}
$, +1/2,\; -1/2,\; +1,\; -1, \ldots$.

Besides, as the~latter condition is satisfied,
$P(\theta )$  takes different forms depending on
the $(j - \lambda)$-correlation:
$$
P(\theta ) \; = \; (1 + \cos \theta )^{j+\lambda } \;
  (1 - \cos \theta )^{j - \lambda } \; = \;
P^{2j}(\cos \theta ),\qquad if\qquad j = \mid \lambda \mid,
\mid \lambda \mid +1,...
$$
or
$$
P(\theta ) \; = \;
{ P^{2j+1}(\cos \theta ) \over \sin \theta }, \qquad if \qquad
 j = \mid \lambda \mid +1/2, \mid \lambda \mid +3/2,...
$$

\noindent so that the second necessary condition  resulting from
the~Pauli criterion  is

2.  {\em given } $\lambda$ {\em according to 1.,
    the number j is allowed to take values}
  $j = \mid \lambda \mid, \mid \lambda \mid +1,...$

Hereafter, these two conditions: 1   and  2 will  be termed  as  the~first
and   respectively   the~second   Pauli consequences.
It should be noted  that  the~angular  variable $\phi $  is  not
affected (charged) by this Pauli condition; in other words, it is
effectively eliminated out of  this  criterion,  but
a~variable that  worked above is the~$\theta$. Significantly,  in
the contrast to this, the well-known procedure [ ] of deriving  the
Dirac  quantization  condition  from investigating continuity
properties of quantum mechanical wave functions,  such
a~working variable is the $\phi $.

If the~first and second Pauli consequences fail, then we face
rather unpleasant mathema\-ti\-cal and physical problems\footnote{A reader  is
referred  to  the Pauli article [65] for more detail about
those peculiarities.}. As a~simple illustration,  we  may  indicate
the~familiar case  when $\lambda= 0$; if in those circumstances,
 the~second  Pauli  condition has failed, then we face the~integer and
half-integer  values  of the orbital angular momentum number
$l = 0, 1/2, 1, 3/2,\ldots\;$
As regards  the Dirac  electron  with  the  components  of  the  total
angular momentum in the form [65]
$$
J_{1} = ( \; l_{1} + \lambda\; { \cos \phi \over \sin  \theta }\; \Sigma _{3}\;),\qquad
J_{2} =(\; l_{2}  + \lambda \; { \sin \phi \over \sin  \theta } \; \Sigma _{3}\;) ,\qquad
 J_{3} = l_{3}
$$

\noindent we have to employ the above Pauli  criterion  in  the  constituent
form owing to $\lambda $  changed into $\Sigma _{3}$
$$
\Sigma _{3} =
\left ( \begin{array}{cccc}
               +1/2 &   0 &  0  &  0 \\
                0   &-1/2 &  0  &  0 \\
                0   &   0 &+1/2 &  0 \\
                0   &   0 &  0  &-1/2
                           \end{array} \right )     .
$$

\noindent Ultimately, we obtain the allowable set $J = 1/2, 3/2, \ldots$.

A~fact of primary  interest  to  us  is  that  the  functions
$\Phi ^{\lambda }_{jm}(\theta, \phi )$ constructed  above  relate
directly  to  the  well-known Wigner $D$-functions (bellow we will use
the notation  according  to [66]):
$$
\Phi ^{\lambda }_{jm}(\theta , \phi ) \; =  \;
(-1)^{j-m} \; D^{j}_{-m, \lambda}(\phi, \theta, 0)
\eqno(2.6)
$$

\noindent Because of the detailed development of $D$-function  theory, this
relation (2.6) will be of great importance in our further work.
Closing this paragraph, we draw attention to   that
the  Pauli  criterion  (here $\Phi ^{\lambda }_{j,-j}(\theta ,\phi )$
denotes  a  spherically symmetrical wave function):
$
J_{-} \Phi _{j,-j}(t,r,\theta ,\phi )\; =\; 0
$
affords a condition that is invariant relative to possible  gauge
transformations. The function $\Phi _{j,m}(t,r,\theta ,\phi )$
may be subjected  to any gauge transformation. But if  all the~components
 $J_{i}$ vary  in a~corresponding way too, then the
Pauli  condition provides the~same result on $J$-quantization. In contrast
to this, the common  requirement to be a~single-valued function of spatial
points is often applied to producing a~criterion on selection  of
allowable wave functions in quantum mechanics, in general, is  not
invariant under gauge transformations and can easily be destroyed
by suitable gauge one.

\subsection*{3.  Electron in a spherically symmetric
                 gravitational field  \\ and Wigner $D$-functions}

Below we  review briefly  some  relevant facts about the TWFI
tetrad formalism. In  the presence of an  external  gravitational
field, the starting Dirac equation
$$
(\; i \gamma ^{a} \partial _{a} \; -
\;m \; ) \Psi (x)\; = \; 0
$$

\noindent is generalized into  [55-63]
$$
[\; i \gamma ^{\alpha }(x) ( \partial _{\alpha } \; + \;
\Gamma _{\alpha }(x) \; ) \; - \; m \; ] \; \Psi (x) \; = \; 0
\eqno(3.1)
$$

\noindent where $\gamma ^{\alpha }(x) \; = \;
\gamma ^{a} e^{\alpha }_{(a)}(x)   , \;
e^{\alpha }_{(a)}(x)$  is a~tetrad; $\Gamma _{\alpha }(x) \; = \;
 {1 \over 2} \sigma ^{ab} \; e^{\beta }_{(a)} \;
\nabla _{\alpha } (e^{\alpha } _{(b)\beta })$   is the bispinor connection;
$\nabla _{\alpha }$ is  the
covariant derivative symbol. In the spinor basis
$$
\psi (x) \; =\;
\left ( \begin{array}{c}
                       \xi(x)  \\     \eta (x)
\end{array} \right ) , \qquad
\xi(x) \; = \; \left ( \begin{array}{c}
                 \xi ^{1}  \\  \xi ^{2}
\end{array} \right ) , \qquad
\eta (x) =
\left ( \begin{array}{c}
            \eta_{\dot{1}} \\ \eta_{\dot{2}}
\end{array} \right ) ,
$$
$$
\gamma ^{a} =
\left ( \begin{array}{cc}
               0   &  \bar{\sigma}^{a} \\
            \sigma ^{a}  &  0
\end{array} \right ) , \qquad
\sigma ^{a} = (I, \; + \sigma ^{k}) , \;\;
 \bar{\sigma}^{a} = (I, \; -\sigma ^{k})
$$

\noindent where ( $\sigma ^{k}$ are the two-row Pauli spin matrices;
 $k =  1, 2, 3$)
we  have  two equations
$$
i \sigma ^{\alpha }(x) \; ( \partial_{\alpha } \; + \;
\Sigma _{\alpha }(x) )\; \xi (x) \; = \; m \; \eta (x) ,
\eqno(3.2a)
$$
$$
i\; \bar{ \sigma ^{\alpha}}(x) \; ( \partial_{\alpha} \; + \;
\bar{\Sigma} _{\alpha }(x) ) \; \eta (x) \;  =  \; m \;   \xi (x)
\eqno(3.2b)
$$

\noindent the symbols $\sigma ^{\alpha }(x), \bar{\sigma} ^{\alpha }(x),
 \Sigma _{\alpha }(x), \bar{\Sigma} _{\alpha }(x)$  denote respectively
$$
\sigma ^{\alpha }(x)  =  \sigma ^{a}\; e^{\alpha }_{(a)}(x) \; , \qquad
\bar{\sigma} ^{\alpha }(x)  =  \bar{\sigma}^{a} \; e^{\alpha }_{(a)}(x)\; ,
$$
$$
\Sigma _{\alpha }(x)  = {1 \over 2} \Sigma ^{ab} e^{\beta }_{(a)}
\nabla _{\alpha }(e_{(b)\beta }) , \qquad
\bar{\Sigma} _{\alpha }(x)  = {1 \over 2}
\bar{\Sigma} ^{ab}e^{\beta }_{(x)} \nabla _{\alpha }(e_{(b)\beta } ) ,
$$
$$
\Sigma ^{ab} = {1 \over 4}
(\bar{\sigma} ^{a} \sigma ^{b}  -  \bar{\sigma} ^{b} \sigma ^{a}) ,   \qquad
\bar{\Sigma} ^{ab} = {1 \over 4} (\sigma ^{a} \bar{\sigma} ^{b}  -
\sigma ^{b} \bar{\sigma} ^{a} )   .
$$

\noindent Setting $m$ equal to zero, we obtain the~Weyl equations for neutrino
$\eta (x)$  and anti-neutrino $\xi (x)$, or Dirac's equation for a~massless
particle.

The form of equations (3.1), (3.2) implies   quite
definite their symmetry properties. It is common, considering the
Dirac equation in the same  space-time,  to  use  some   different
tetrads $e^{\beta }_{(a)}(x)$ and $e^{'\beta}_{b}(x)$, so that we have
the~equation  (3.1)
and analogous one with a~new tetrad mark. In other words, together
with (3.1) there exists an equation on $\Psi'(x)$  where the quantities
$\gamma{'\alpha }(x)$ and $\Gamma'_{\alpha}(x)$, in comparison with
$\gamma^{\alpha}(x)$ and $\Gamma_{\alpha}(x)$, are  based
on another tetrad $e^{'\beta }_{b)}(x)$ related to $e^{\beta }_{(a)}(x)$
 through  some  local Lorentz matrix
$$
e_{(b)}^{'\beta} (x) \;  = \; L^{\;\;a}_{b}(x) \; e^{\beta }_{(a)}(x) \; .
\eqno(3.3a)
$$

\noindent It may be shown that these two Dirac equations on  functions
$\Psi (x)$ and $\Psi'(x)$ are related to each other by a~quite definite
bispinor transformation
$$
\xi'(x) \; = \; '(k(x)) \; \xi (x) \; , \qquad
\eta'(x) \; = \; '^{+} (\bar{k}(x))\; \eta (x) .
\eqno(3.3b)
$$

\noindent Here, $B(k(x)) = \sigma ^{a} k_{a}(x)$ is a~local matrix from
the $SL(2.C)$ group; 4-vector $k_{a}$ is the well-known parametre
on this group [74,75].  The
matrix $L^{\;\;a}_{b}(x)$ from (3.3a)  can be expressed  as  a~function  of
arguments $k_{a}(x)$  and $k^{*}_{a}(x)$:
$$
L^{\;\;a}_{b}(k, k^{*}) \; = \; \bar{\delta}^{c}_{b}  \;\;
[ \; - \delta ^{a}_{c} \; k^{n} \; k^{*}_{n} \; + \;
k_{c} \; k ^{a*} \; + \;
k^{*}_{c} \; k^{a} \; + \;
i \; \epsilon ^{anm}_{c} \; k_{n} \; k^{*}_{m} \; ]
\eqno(3.3c)
$$

\noindent where $\bar{\delta} ^{c}_{b}$ is a~special Cronecker's symbol
$$
\bar{ \delta} ^{c}_{b} =
\left \{ \begin{array}{l}
                0 ,\; \;  if \;\;  c \neq  b ;\\
                +1 ,\;\;  if \;\;  c = b = 0  ; \\
                -1 ,\;\;  if \;\;  c = b = 1,2,3
\end{array} \right.
$$

It is normal practice that some different  tetrads  are
used at examining the~Dirac equation on the~background of a~given
Rimaniann space-time. If there is  a~need  for  analysis  of
the~correlation between  solutions in such distinct tetrads,  then  it
is important  to know how to  calculate  the~corresponding  gauge
transformations over the spinor wave functions.  First,  the~need
for taking into account such a~gauge transformation was especially
emphasized  by Fock V.I.  [57]. The first who were interested in
explicit expressions for such spinor matrices, were
E.~Schr\"odinger [64] and  W.~Pauli  [65]. Thus, Schr\"odinger
found the~matrix  relating  spinor  wave  functions  in Cartesian
and  spherical tetrads:
$$
x^{\alpha } = (x^{0},x^{1},x^{2},x^{3}),\;\;
dS^{2} = [(dx^{0})^{2} - (dx^{1})^{2} - (dx^{2})^{2} - (dx^{3})^{2} ],\;\;
e^{\alpha }_{(a)} (x) = \delta ^{\alpha }_{a}
\eqno(3.4a)
$$

\noindent and
$$
x^{'\alpha } = ( t , r , \theta  , \phi  ) , \;\; dS^{2} =
[\; dt^{2} \; - \; dr^{2} \; - \;  r^{2} \;
 (d\theta ^{2} \; + \; \sin ^{2}\theta  d\phi ^{2})\; ],
$$
$$
e^{\alpha'}_{(0)} = ( 1 , 0 , 0 , 0 ) , \qquad
e^{\alpha '}_{(1)} = ( 0 , 0 ,1/r, 0 ) ,
$$
$$
e^{\alpha '}_{(2)} = ( 0 ,0 , 0 , {1 \over r \;  \sin \theta}), \qquad
e^{\alpha '}_{(3)} = ( 0 , 1 , 0 , 0 )
\eqno(3.4b)
$$

\noindent the relevant matrix is (where  $\;\vec{c}\;$ is the Gibbs parametre
on the group $S0(3.R)$; see for more details in [75])
$$
B = \; \pm \;
\left ( \begin{array}{cc}
       \cos \theta/2 \; e^{i\phi /2}   & \sin \theta/2  \; e^{-i\phi /2} \\
       -\sin \theta /2 \; e^{i\phi /2}   & \cos \theta /2 \;  e^{-i\phi /2}
\end{array} \right )
 \; \equiv \;  B(\vec{c}) = \; \pm \; { \; I \; - \;i \; \vec{\sigma } \; \vec{c}
\over
\sqrt{1 - (\vec{c})^{2}}}
$$

This basis of spherical tetrad will play a~substantial  role
in our farther work. Just one (the spherical tetrad's  basis)  was
used with great efficiency by Pauli [65]   when   investigating
the~problem of allowed spherically symmetrical  wave  functions  in
quantum mechanics.  Now, let us  reexamine  the~problem  of  free
electron in the external spherically symmetric gravitational field
(see also in [68-70] about the~manner of working on this in the frame of
the~so-called light tetrad or Newman-Penrose's formalism),
but centering upon  some facts  which  will  be  of  great
importance at extending that method on  an~electron-monopole
system.

In particular,  we  consider briefly a~question of separating the~angular variables in
the Dirac equation on the background of  a~spherically  symmetric
Rimanian space-time. As a~starting point we take a~flat space-time
model, so that an original equation (3.1) being specified for  the
spheric tetrad (see (3.4b)) takes on the form
$$
\left [ \;i \;
\gamma ^{0} \; \partial_{t} \;   + \;
i \;  ( \gamma ^{3} \; \partial_{r} \; + \; { \gamma ^{1} \; \sigma ^{31} \; + \;
\gamma ^{2} \; \sigma ^{32} \over  r } ) \;  +
{1 \over r} \; \Sigma_{\theta \phi} \;  - \; m \;
\right ]\;  \Psi (x) \; = \; 0
\eqno(3.5a)
$$

\noindent where
$$
\Sigma _{\theta ,\phi } \; = \; \left [
 i\; \gamma ^{1} \partial _{\theta} \;+\;
\gamma ^{2} \;  {\;  i \partial _{\phi} \; + \;
 i\; \sigma ^{12} \over \sin \theta } \; \right ]\; .
\eqno(3.5b)
$$

We  specialize   the~electronic   wave   function   through
substitution   (Wigner functions are designated by
$D^{j}_{-m,\sigma }(\phi ,\theta ,0)  \equiv  D_{\sigma}$)
$$
\Psi _{\epsilon jm}(x) \;  = \; {e^{-i\epsilon t} \over r} \;
\left ( \begin{array}{l}
        f_{1}(r) \; D_{-1/2} \\ f_{2}(r) \; D_{+1/2}  \\
        f_{3}(r) \; D_{-1/2} \\ f_{4}(r) \; D_{+1/2}
\end{array} \right ).
\eqno(3.6)
$$

\noindent Using recursive formulas (see in  [66] )
$$
\partial_{\theta} \; D_{+1/2} \; = \; ( a\; D_{-1/2}  - b \; D_{+3/2} ) ,\;\;
{- m - 1/2 \;  \cos \theta  \over  \sin \theta } \; D_{+1/2}\; =\;
(- a \;  D_{-1/2} - b \;  D_{+3/2} ) ,
$$
$$
\partial_{\theta} \;  D_{-1/2} \; = \; ( b \; D_{-3/2} - a \; D_{+1/2} ) ,\;\;\;
{- m + 1/2 \; \cos \theta \over \sin \theta}  \;  D_{-1/2} \; = \;
 (- b \; D_{-3/2}  - a \;  D_{+1/2} ) \; ,
$$

\noindent where   $a = (j + 1)/2$  and
$b  = {1\over 2} \;\sqrt{(j-1/2)(j+3/2)}$,  we find ( $ \nu  = (j + 1/2)/2$  )
$$
 \Sigma _{\theta ,\phi } \; \Psi _{\epsilon jm}(x) \;  = \; i\; \nu \;
{e^{-i\epsilon t } \over r} \;
\left ( \begin{array}{r}
        - \; f_{4}(r) \; D_{-1/2}  \\  + \; f_{3}(r) \; D_{+1/2} \\
        + \; f_{2}(r) \; D_{-1/2}  \\  - \; f_{1}(r) \; D_{+1/2}
\end{array} \right )
\eqno(3.7)
$$

\noindent  further one gets the following set of radial equations
$$
\epsilon   f_{3}   -  i  {d \over dr}  f_{3}   -
i {\nu \over r}  f_{4}  -  m  f_{1} =   0  , \qquad
\epsilon   f_{4}   +  i  {d \over dr} f_{4}   +
i {\nu \over r}  f_{3}  -  m  f_{2} =   0     ,
$$
$$
\epsilon   f_{1}   +  i  {d \over dr}  f_{1}  +
i {\nu \over r}  f_{2}  -  m  f_{3} =   0 , \qquad
\epsilon   f_{2}   -  i  {d \over dr} f_{2}   -
i {\nu \over r}  f_{1}  -  m  f_{4} =   0  .
\eqno(3.8)
$$

The usual $P$-reflection symmetry operator  in  the  Cartesian
tetrad basis is $\hat{\Pi}_{C.} \; = \; i \gamma ^{0} \otimes
\hat{P}$, or in a more detailed form
 $$
  \hat{\Pi}_{C.}  = \left (
\begin{array}{cccc} 0 &  0 &  i &   0  \\ 0 &  0 &  0 &   i  \\ i &
          0 &  0 &   0  \\ 0 &  i &  0 &   0 \end{array} \right )
\; \otimes  \; \hat{P} \; ,
\qquad \hat{P} (\theta , \phi ) = (\pi  -
\theta, \; \phi+ \pi )
 $$

\noindent being subjected to translation into the spherical one
$
\hat{\Pi}_{sph.} = S(\theta ,\phi )  \hat{\Pi}_{C.}
 S^{-1}(\theta ,\phi )$
gives us the~result
$$
\hat{\Pi}_{sph.} \; \; =
\left ( \begin{array}{cccc}
0 &  0 &  0 & -1   \\
0 &  0 & -1 &  0   \\
0 &  -1&  0 &  0   \\
-1&  0 &  0 &  0
\end{array} \right )
\; \otimes  \; \hat{P} \; .
\eqno(3.9)
$$

\noindent From the equation on proper values
$$
\hat{\Pi}_{sph.}\; \Psi _{jm}
= \; \Pi \; \Psi _{jm}\;\;  when \;\;
(\; \hat{P} \; D^{j}_{-m,\sigma } (\phi ,\theta ,0)  =
(-1)^{j} \; D^{j}_{-m,-\sigma} (\phi ,\theta ,0)\; )
$$

\noindent we get
$$
\Pi = \; \delta \;  (-1)^{j+1} , \;\; \delta  = \pm 1 : \qquad
f_{4} = \; \delta \;  f_{1} , \qquad  f_{3} = \;\delta \; f_{2}
\eqno(3.10)
$$

\noindent so that $\Psi _{\epsilon jm\delta }(x)$ is
$$
\Psi (x)_{\epsilon jm\delta } \; = \; {e^{-i\epsilon t} \over  r } \;
\left ( \begin{array}{r}
     f_{1}(r) \; D_{-1/2}(\theta ,\phi ,0) \\
     f_{2}(r) \; D_{+1/2}(\theta ,\phi ,0) \\
\delta \; f_{2}(r) \; D_{-1/2}(\theta ,\phi ,0)   \\
\delta \; f_{1}(r) \; D_{+1/2}(\theta ,\phi ,0)
\end{array} \right )  .
\eqno(3.11)
$$

\noindent Noting (3.10), we simplify the~system  (3.8); it is  reduced  to
$$
({d \over dr} \;+\;{\nu \over r}\;) \; f \; + \; ( \epsilon  \;+ \;
 \delta \; m )\; g \; = \;0 \;  ,  \qquad
({d \over dr} \; - \;{\nu \over r}\;)\; g  \;- \; ( \epsilon \; - \;
 \delta\;  m )\; f\; =\; 0
\eqno(3.12)
$$

\noindent where instead of $f_{1}$  and $f_{2}$  we have employed their
linear combinations
$$
f \; = \; {f_{1} + f_{2} \over \sqrt{2}} , \qquad
g \; = \; {f_{1} - f_{2} \over i \sqrt{2}} \; .
$$

It should be useful to notice that the~above  simplification
$( \Psi _{\epsilon jm} \rightarrow  \Psi _{\epsilon jm\delta } )$  can
also  be   obtained   through   the~diagonalization of the
operator  $\hat{K}$ (see in [65]):
$$
\hat{K} \; = \;  - \gamma ^{0} \gamma ^{3} \; \Sigma _{\theta ,\phi } =
\gamma ^{0}\gamma ^{3} \left [\;
 \gamma ^{1} \; (\partial _{\theta } + 1/2) +
 {\gamma ^{2} \over \sin \theta } \; \partial_{\phi} \;\right ] \; .
\eqno(3.13a)
$$

\noindent Actually, from  $\hat{K} \; \Psi _{\epsilon jm} (x) =
K \; \Psi _{\epsilon jm}$ we produce
$$
K = - \delta \; (j+1/2) \;\;  ,\;\;  \delta \; = \pm  1 :\qquad
f_{4} = \; \delta \; f_{1} ,\qquad f_{3} = \; \delta \;  f_{2}  \; .
\eqno(3.13b)
$$

Everything established above for the flat space-time  model  can
be readily generalized  into  an arbitrary  curved  space-time  with
a spherically symmetrical metric $g_{\alpha \beta }(x)$:
$$
dS^{2} = [\; e^{\nu }\; (dt)^{2} \; - \; e^{\mu } \; (dr)^{2} \; - \;
r^{2} \; ((d\theta )^{2} + \sin ^{2}\theta  (d\phi )^{2})\; ]\; ,
\eqno(3.14a)
$$

\noindent and  its naturally coresponding diagonal tetrad
 $e^{\alpha }_{(a)}(x)$:
$$
e^{\beta }_{(0)} = ( e^{-\nu /2}, 0 , 0 , 0) \; , \qquad
e^{\beta }_{(3)} = (0 , e^{-\mu /2}, 0 , 0 ) \; ,
$$
$$
e^{\beta }_{(1)} = (0, 0 , {1 \over r} , 0) \; , \qquad
e^{\beta }_{(2)} = (0, 0 , 0 , {1 \over r \sin \theta } ) \; .
\eqno(3.14b)
$$

\noindent The general covariant Dirac equation can be specified, according to
[57], for an~arbitrary diagonal tetrad as follows
$$
\left [\;i\; \gamma ^{a}\; ( e^{\beta }_{(a)} \; \partial_{\beta } \;
+ \;{1 \over 2}\; e^{\beta }_{(a);\beta } ) \;  - \;  m\right ] \; \Psi(x) = 0
\eqno(3.15a)
$$

\noindent where the $e^{\beta }_{(a);\beta }$ can be computed by means of
$$
e^{\beta }_{(a);\beta } \; = \; { 1 \over \sqrt{ - \det g }}\; {\partial \over
\partial x^{\beta}}  \;  \sqrt{-\det g } \; e^{\beta }_{(a)}\; .
\eqno(3.15b)
$$

\noindent So, for the function $\Phi (x)$ defined by
$$
\Psi (t,r,\theta ,\phi )\; =\; \exp (-{1 \over 4} (\nu +\mu )) \;\; {1 \over r}\; \;
\Phi  (t,r,\theta ,\phi )
\eqno(3.16a)
$$

\noindent we  produce the equation
$$
\left [\; i \; \gamma ^{0} \; e^{-\nu /2} \; \partial_{t} \; +\;
i \; \gamma ^{3}\; e^{-\mu /2} \; \partial_{r} \; + \;
 {1 \over r} \; \Sigma _{\theta ,\phi }\;  -\;  m \;
\right ] \; \Phi  (t,r,\theta ,\phi )= 0\; .
\eqno(3.16b)
$$

On comparing (3.16b) with (3.5a), it follows immediately                                       that
all the calculations carried out above for the flat space-time case are
still valid only with some evident modifications. Thus,
$$
\Phi _{jm\delta }(x)  =
\left ( \begin{array}{rrrr}
  f_{1}(r,t) \; D_{-1/2}(\theta ,\phi ,0) \\
  f_{2}(r,t) \; D_{+1/2}(\theta ,\phi ,0) \\
  \delta \;  f_{2}(r,t) \; D_{-1/2}(\theta ,\phi ,0)   \\
  \delta \;  f_{1}(r,t) \; D_{+1/2}(\theta ,\phi ,0)
\end{array} \right )
\eqno(3.17a)
$$

\noindent and instead of (3.12) now we find
$$
( e^{-\mu /2} {d \over dr} + {\nu \over r} ) f \; + \;
( ie^{-\nu /2} \; \partial_{t} \; +\;  \delta  \; m )\;  g \; = 0\; ,
$$
$$
( e^{-\mu /2} {d \over dr} - {\nu \over r} ) g \; - \;
 ( ie^{-\nu /2} \; \partial_{t} \; -\; \delta \;  m ; f\;   = 0 \; .
\eqno(3.17b)
$$

\subsection*{4.   Electronic wave functions
                  in the external monopole field.}

In  the~literature,  the~electron-monopole   problem   has
attracted a~lot of attention. In particular,  the~various
properties  of  occurring  so-called   monopole   harmonics   were
investigated in great detail. Here, we are going to look into this
problem in the~context of generalized  Pauli-Schr\"odinger  formalism
reviewed in Sections 2-3. At this we seek to maintain as  close
connection as possible with the preceeding formalism.

For our further purpose  it  will  be  convenient  to  use  a
monopole Abelian potential in the Scwinger's form:
$$
A^{a}(x) = (A^{0}, \; A^{i}) = \left ( 0 \; , \; g\;
 {(\vec{r} \times \vec{n})\;(\vec{r} \; \vec{n}) \over
r \; (r^{2} - (\vec{r} \; \vec{n})^{2}) } \right )
\eqno(4.1a)
$$

\noindent after translating  the $A_{\alpha }$  to  the~spherical
coordinates  and specifying $\vec{n} = (0, 0 , 1 )$ ,  we get
$$
A_{0} = 0 , \;\; A_{r} = 0 , \;\; A_{\theta } = 0\; , \qquad
A_{\phi } = g\; \cos \theta  \; .
\eqno(4.1b)
$$

\noindent Correspondingly,  the~Dirac  equation  in  this   electromagnetic
potential takes the form
$$
\left [ i \gamma ^{0} \partial _{t} + i \gamma ^{3} (\partial _{r} +
{1 \over r}) + {1 \over r}\; \Sigma ^{k}_{\theta ,\phi } \; -\;
 mc/\hbar  \right ]\; \Psi (x) = 0
\eqno(4.2a)
$$

\noindent where
$$
\Sigma ^{k}_{\theta ,\phi } \; = \; \left [ \;
 i \gamma ^{1} \partial _{\theta} \;  + \;
\gamma ^{2} \;  { i \partial _{\phi } + (i\sigma ^{12} - k ) \;
 \cos \theta \over  \sin  \theta}  \; \right ]
\eqno(4.2b)
$$

\noindent and $k \equiv  eg/hc$.   As readily verified, the~wave  operator
  in  (4.2a ) commutes with the~following three ones
$$
J^{k}_{1} = \left [ \; l_{1} + {(i\sigma ^{12} - k)
 \cos \phi  \over \sin \theta } \;\right ]\; ,  \;\;
J^{k}_{2} = \left [ \; l_{2} + {(i\sigma ^{12} - k)
\sin \phi  \over \sin \theta } \;\right ]\; , \;\;\;
\qquad  J^{k}_{3} = l_{3}
\eqno(4.3a)
$$

\noindent which  in  turn    obey   the~$SU(2)$   Lie   algebra.   Clearly,
this    monopole     situation     come  entirely  under
the~Schwinger-Pauli approach, so that our  further  work  will  be
a~matter of simple (quite elementary) calculations.

Thus, corresponding to diagonalization of the
 $\vec{J}^{2}_{k}$ and $J^{k}_{3}$,
the~function $\Psi$' is to be initially taken as
($D_{\sigma } \equiv D^{j}_{-m,\sigma }(\phi ,\theta ,0)$)
$$
\Psi ^{k}_{\epsilon jm} (t,r,\theta ,\phi ) = {e^{-i\epsilon t} \over  r} \;
\left ( \begin{array}{r}
       f_{1} \; D_{k-1/2}   \\   f_{2} \; D_{k+1/2}   \\
       f_{3} \; D_{k-1/2}   \\   f_{4} \; D_{k+1/2}
\end{array} \right )\; .
\eqno(4.3b)
$$

\noindent Further, noting recursive relations  [66]
$$
\partial_{\theta} \;  D_{k+1/2} = (+ a \; D_{k-1/2} - b \; D_{k+3/2} )
\; \;,
$$
$$
{-m -(k+1/2) \cos \theta \over \sin \theta } \; D_{k+1/2} =
(- a \; D_{k-1/2} - b \; D_{k+3/2} )  \;\; ,
$$
$$
\partial_{\theta} \;  D_{k-1/2} = (+ c \; D_{k-3/2} - a \; D_{k+1/2} )
\;\; ,
$$
$$
{-m -(k-1/2)\cos \theta \over \sin \theta }\; D_{k-1/2} =
(- c \; D_{k-3/2} - a \; D_{k+1/2} )
$$

\noindent where
$$
a = {1 \over 2} \sqrt{(j + 1/2)^{2} - k^{2}} \; ,\;\; \;
b = {1 \over 2} \sqrt{(j - k - 1/2)(j + k + 3/2)}\; , \;\;\;
$$
$$ c = {1 \over 2} \sqrt{(j + k - 1/2)(j - k + 3/2)}
$$

\noindent we find how the $\Sigma ^{k}_{\theta ,\phi }$  acts on $\Psi $:
$$
\Sigma ^{k}_{\theta ,\phi } \; \Psi ^{k}_{\epsilon jm} =
 i \;\; \sqrt{(j + 1/2)^{2} - k^{2}} \;\; {e^{-i\epsilon t} \over r} \;
\left ( \begin{array}{r}
  - f_{4} \; D_{k-1/2}  \\  + f_{3} \; D_{k+1/2} \\
  + f_{2} \; D_{k-1/2}  \\  - f_{1} \; D_{k+1/2}
\end{array} \right )
\eqno(4.4)
$$

\noindent hereafter the factor $\sqrt{(j + 1/2)^{2}- k^{2}}$
 will be denoted by $\nu $. For the radial $f_{i}(r)$  we establish
$$
\epsilon \; f_{3} \; - \; i \; {d\over dr} \; f_{3}  \;- \;i\; {\nu \over r}\; f_{4}\;
 - \; m \; f_{1} = 0 \; ,
\qquad
\epsilon \; f_{4} \; + \; i\;  {d \over dr}\;  f_{4}\; + \;
 i \;{\nu \over r} \; f_{3} \; - \; m \; f_{2} = 0 \; ,
$$
$$
\epsilon \; f_{1} \; +\; i\; {d \over dr} \; f_{1} \;  + \;
i \; {\nu \over r} \; f_{2} \;- \;m\; f_{3} = 0 \; ,
\qquad
\epsilon \; f_{2}\; - \;i\; {d \over dr}\; f_{2}\; - \;
i \;{\nu \over r}\; f_{1}\; -\; m \;f_{4} = 0 \; .
\eqno(4.5)
$$

As evidenced by analogy with preceding  Sec.3   and  also  on
direct  calculation,  else  one  operator  can  be  simultaneously
diagonalized together with $\{ i \; \partial _{t} , \vec{J}^{2}_{k},
J^{k}_{3} \}$,  namely, a~generalized Dirac operator
$$
\hat{K} ^{k} \; = - \; i \; \gamma^{0} \; \gamma ^{3} \;
\Sigma ^{k}_{\theta ,\phi }   \; .
\eqno(4.6a)
$$

\noindent From the  equation $\hat{K}^{k} \Psi _{\epsilon jm}  =
K \; \Psi _{\epsilon jm}$    we  can  produce  two
possible values for this~$K$  and the corresponding limitations on $f_{i}(r)$:
$$
K = - \delta \;  \sqrt{(j + 1/2)^{2}- k^{2}} \; : \qquad
f_{4} = \delta \;  f_{1} \; , \;\;\; f_{3} = \delta \; f_{2}
\eqno(4.6b)
$$

\noindent and in a~consequence of this, the~system  (4.5) is reduced to
$$
({d \over dr} + {\nu \over r}) f \; + \; (\epsilon  + \delta\;  m )\;  g = 0\; ,
\;\;\;\;
({d \over dr} - {\nu \over r}) g \; - \; (\epsilon  - \delta\;  m ) \; f = 0\;.
\eqno(4.7)
$$

\noindent On direct comparing  (4.7)  with analogous system in  Sec.3,
we can conclude that these systems are formally similar apart  from
the difference  between $\nu  = j +1/2$  and $\nu  =
 \sqrt{(j + 1/2)^{2}- k^{2}}$.

Now let us pass over to quantization of  $k = eg/hc$  and $J$.  As
a~direct result  from   the~first  Pauli condition (2.5a) we derive
that $$ {eg \over hc} = \pm 1/2 , \; \pm 1, \; \pm 3/2, \ldots
\eqno(4.8a)
$$

\noindent which coincides with the Dirac's  quantization,  and  from
the~second Pauli consequence it follows immediately that
$$
k = {eg \over \hbar c}  = \pm  1/2, \pm  1, \pm  3/2,\ldots
\;\;\; and \;\;\;
j = \mid k \mid  -1/2, \mid k \mid +1/2, \mid k \mid +3/2,\ldots
\eqno(4.8b)
$$

The case of minimal allowable value $j_{min.}= \mid k \mid - 1/2$  must  be
separated out and looked into in a special way.  For example, let
$k = +1/2$, then to the minimal value $j = 0$ there corresponds a~wave
function in terms of solely $(t,r)$-dependent quantities
$$
\Psi ^{(j=0)}_{k = +1/2}(x) = { e^{-i\epsilon t} \over  r}
\left ( \begin{array}{l}
           f_{1}(r)  \\   0  \\  f_{3}(r)  \\  0
\end{array} \right )   \; .
\eqno(4.9a)
$$

\noindent At $k = - 1/2$, in an analogous way,  we have
$$
\Psi ^{(j =0)}_{k = -1/2}(x) =
{e^{-i\epsilon t} \over  r}
\left ( \begin{array}{l}
   0  \\  f_{2}(r)  \\   0   \\  f_{4}(r)
\end{array} \right )       \;          .
\eqno(4.9b)
$$

\noindent Thus, if $k = \pm  1/2$, then to the minimal alowed values
 $J_{\min }$
there correspond the function substitutions which do not depend at
all on the angular variables $(\theta ,\phi )$; at
 this point there exists some
formal analogy between  these  electron-monopole  states  and
$S$-states ( with $l = 0 $) for a~boson field of spin zero:
$\Phi _{l=0} = \Phi (r,t)$. However, it would be unwise to attach too much
significance
to this formal coincidence  because such a~$(\theta ,\phi )$-independence
of $(e-g)$-states  is  not  a~fact  invariant   under   tetrad   gauge
transformations. In contrast, the relation below (let $k = +1/2)$
$$
\Sigma^{+1/2}_{\theta ,\phi } \; \Psi ^{(j=0)}_{k=+1/2} (x) \; = \;
\gamma ^{2} \; \; \cot \theta \; ( i \sigma ^{12} - 1/2 ) \;\;
\Psi ^{(j =0)} _{k=+1/2} \equiv  0
\eqno(4.10a)
$$

\noindent is invariant under any gauge transformations. The identity (4.10a)
holds because all the zeros in the
$\Psi ^{(j=0)}_{k=+1/2}$ are adjusted to the non-zeros in
$( i \sigma ^{12} - 1/2 )$;  and
conversely,  the  non-vanishing  constituents  in $\Psi ^{(j=0)}_{k=+1/2}$
are canceled out by zeros in $( i \sigma ^{12}- 1/2 )$.
Correspondingly, the~matter equation (4.2a) takes on the form
$$
\left [\;  i \; \gamma ^{0} \; \partial_{t} \; +
\; i\; \gamma ^{3} \; (\partial_{r}
\; + \;  {1 \over r}\;  )\; - \;  mc/\hbar \; \right ] \; \Psi ^{(j=0)} = 0\; .
\eqno(4.10b)
$$

It is readily  verified  that  both (4.9a)  and (4.9b)
representations are
directly extended to $(e-g)$-states  with $j = j_{\min }$ at all  the other
$k =\pm 1, \pm 3/2, \ldots $.  Indeed,
$$
k= +1, +3/2, +2,\ldots : \qquad
\Psi ^{k > 0} _{j_{min.}} (x) = {e^{-i\epsilon t} \over r}
\left ( \begin{array}{l}
   f_{1}(r) \; D_{k-1/2}  \\  0  \\  f_{3}(r) \;  D_{k-1/2} \\  0
\end{array} \right ) \; ;
\eqno(4.11a)
$$
$$
k = -1, -3/2,-2,\ldots : \qquad
\Psi ^{k<0} _{j_{min.}} (x) = { e^{-i\epsilon t} \over  r}
\left ( \begin{array}{l}
    0    \\   f_{2}(r) \; D_{k+1/2}  \\  0  \\ f_{4}(r) \; D_{k+1/2}
\end{array} \right )
\eqno(4.11b)
$$

\noindent and, as can be shown,  the relation
$\Sigma _{\theta ,\phi } \Psi _{j_{\min }} = 0 $ still  holds.
For instance, let us consider in more detail  the case of positive $k$.
Using the recursive relations [66]
$$
\partial _{\theta } \; D_{k-1/2} =
 { 1 \over 2} \sqrt{ 2k-1} \; D_{k-3/2}\; , \qquad
{- m - (k-1/2) \cos \theta  \over \sin \theta } \; D_{k - 1/2}  =
 - { 1 \over 2} \sqrt{ 2k -1} \;  D_{k-3/2}\; ,
$$
\noindent we get
$$
i\gamma ^{1} \; \partial _{\theta}
\left ( \begin{array}{c}
       f_{1}(r) \; D_{k-1/2} \\  0  \\  f_{3}(r) \; D_{k-1/2}  \\  0
\end{array} \right ) = {i\over 2} \sqrt{2k-1} \;
\left ( \begin{array}{c}
     0  \\ - f_{3}(r) \; D_{k-3/2}  \\ 0  \\ + f_{1}(r)\; D_{k-3/2}
\end{array} \right ) \; ;
$$
$$
\gamma ^{2}  \; {i\partial _{\phi } + (i\sigma ^{12} - k) \cos \theta \over
\sin \theta} \;
\left ( \begin{array}{c}
      f_{1}(r) \; D_{k-1/2} \\  0  \\  f_{3}(r) \; D_{k-1/2} \\ 0
\end{array} \right ) \; = \;
{i \over 2} \sqrt{2k-1} \;
\left ( \begin{array}{c}
    0 \\ +f_{3}(r) \; D_{k-3/2}  \\ 0  \\ -f_{1}(r) \; D_{k-3/2}
\end{array} \right )
$$

\noindent in a~sequence, the identity
$\Sigma _{\theta ,\phi } \; \Psi _{j_{\min }} \equiv  0$  has been proved.
The  case of negative $k$ can be considered in the same way.

Thus, at every $k$, the $j_{\min }$-state's equation  has  the  same
unique form
$$
\left [ \; i\; \gamma ^{0} \; \partial_{t} \; + \; i\gamma ^{3} \;
(\partial_{r}\; + \; {1 \over r}\; ) \;  - \;  mc/\hbar \;
\right ] \; \Psi _{j_{mi}} = 0
\eqno(4.11c)
$$

\noindent which leads to the radial system
$$
k = +1/2,+1,\ldots : \;\;\;
\epsilon \; f_{3} - i \; { d\over dr}  \; f_{3}  - m \; f_{1} = 0\; , \;\;\;
\epsilon \; f_{1} + i \; { d \over dr} \; f_{1}  - m \; f_{3} = 0 \; ;
\eqno(4.12a)
$$
$$
k = -1/2,-1,\ldots \; : \;\; \;
\epsilon \; f_{4} + i \; { d\over dr}\; f_{4} - m \;f_{2} = 0 \; , \;\;\;
\epsilon \; f_{2} - i \; { d\over dr}\; f_{2} - m \;f_{4} = 0 \; .
\eqno(4.12b)
$$

\noindent These equations are equivalent respectively to
$$
k = + 1/2,+ 1,\ldots \; : \qquad
\left [ {d^{2} \over dr^{2}}  + \epsilon ^{2}  - m^{2}\right ]\; f_{1}  = 0 \;
, \qquad  f_{3} =  { 1 \over m}\left ( \epsilon  +
i { d \over dr }\right ) \; f_{1}     \; ;
\eqno(4.13a)
$$
$$
k = - 1/2, - 1,\ldots \; :  \qquad
\left [{d^{2} \over dr^{2}}  + \epsilon ^{2}  - m^{2}\right ] \; f_{4} = 0\;
 ,\qquad f_{2} = {1 \over m} \left ( \epsilon  + i {d \over dr} \right )
 \; f_{4}
\eqno(4.13b)
$$

\noindent which both end up  with the functions
 $f = \exp  ( \pm  \sqrt{m^{2} - \epsilon ^{2}} \; r )$.
This latter, at $\epsilon \; < \; m$, looks as
$$
\exp  [-\sqrt{m^{2} -\epsilon ^{2}} \; r ]
\eqno(4.13c)
$$

\noindent which seems  to  be appropriate to describe   a~bound   state   in
the~electron-monopole system. It  should  be  amphasised  that
today the $j_{\min }$   bound  state  problem   remains   a~still  yet
question to understand.  In  particular,  the  important  question
faced us is of finding a~physical and mathematical   criterion  on
selecting values for $\epsilon $: whether $\epsilon \; < \; m$  , or
$\epsilon  = m$ , or $\epsilon \; > \; m$;
and what value of $\epsilon $  is to be chosen  after specifying an~interval
above.

Now let us  proceed with studying the  properties  which
stem  from  the $\theta ,\phi $-dependence  of  the  wave   functions.
In particular, we restrict ourselves to  the $P$-parity
problem  in  the  presence  of  the  monopole.  This  problem  was
investigated in some detail in the literature [11-16,76-85],
so our first  step is to particularize some relevant facts in accordance with the
formalism and notation used in the present paper.

As evidenced by straightforward computation,  the  well-known
purely  geometrical  bispinor $P$-reflection  operator  does not
commute with the Hamiltonian $\hat{H}$  under  consideration. The same
conclusion  is also arrived at by  attempt to solve  directly  the
proper value equation
$$
\hat{\Pi}_{sph.} \; \Psi ^{k} _{\epsilon jm} =
 \Pi \; \Psi ^{k}_{\epsilon jm}
$$

\noindent which leads to
$$
(-1)^{j+1} \;
\left ( \begin{array}{l}
   f_{4} \; D_{-k-1/2} \\   f_{3} \; D_{-k+1/2} \\
   f_{2} \; D_{-k-1/2} \\   f_{1} \; D_{-k+1/2}
\end{array} \right ) \; = \;  P \;
\left ( \begin{array}{l}
   f_{1} \; D_{k-1/2}   \\  f_{2} \; D_{k+1/2} \\
   f_{3} \; D_{k-1/2}   \\  f_{4} D_{k+1/2}
\end{array} \right  )
$$

\noindent the  latter  matrix  relation  is  satisfied  only  by  the~trivial
substitution $f_{i}= 0$   for  all  $i$.  The~ matrix  relation above
indicates how a~required discrete transformation can be  constructed
(further we will denote it as $\hat{N}_{sph.}$ )
$$
\hat{N}_{sph.} \;  =    \; \hat{\pi } \otimes
\Pi_{sph.} \otimes  \hat{P}
\eqno(4.14)
$$

\noindent where $\hat{\pi }$  is a~special discrete operator changing
$k ( = eg/hc)$ into $-k: \hat{\pi } \; F( k ) \;  = \; F (- k )$.
Such an operator $\hat{N}_{sph.}$  commutes with $\hat{H}$  and
 $\hat{J}^{k}_{i}$; besides,  from
the equation $\hat{N}_{sph.}\; \Psi ^{k}_{\epsilon jm} =
N \Psi ^{k}_{\epsilon jm}$  it follows
$$
N = \; \delta \; (-1)^{j+1} \; ( \delta  = \pm  1 ) : \qquad
f_{4} = \delta \; f_{1} ,\;\; f_{3} = \delta \; f_{2} \; .
\eqno(4.15a)
$$

\noindent The latter relations are compatible with the~above  radial system
(4.5) and they are transformed   into ( $f(r)$ and $g(r)$  are  already
used combinations from $f_{1}(r)$  and $f_{2}(r)$)
$$
({d \over dr} + {\nu \over r})  f + (\epsilon  + \delta \;  m ) g = 0\; ,
 \qquad
({d \over dr} - {\nu \over r}) g -  (\epsilon  - \delta \;  m ) f = 0
\eqno(4.15b)
$$

\noindent that coincides  with (4.7).

We are to say that everything just  said  about  diagonalizing  the
$\hat{N}_{sph.}$ is applied only to the cases when $j > j_{\min }$.
As  regards the~lower value of $j$, the situation turns out to be very
specific and unexpected. Actually, let $k = + 1/2 $ and $-1/2$  ($j  = 0$);
then we have
$$
\hat{N}_{sph.} \; \Psi ^{(j=0)} = N \; \Psi ^{(j=0)} \;\; \rightarrow \;\;
\left ( \begin{array}{r}
         0  \\  - f_{3} \\  0  \\ - f_{4}
\end{array} \right ) \; = \; N \;
\left ( \begin{array}{r}
          f_{1}\\  0 \\ f_{3} \\  0
\end{array} \right )   \; ;
$$
$$
\hat{N}_{sph.} \; \Psi ^{(j=0)} = N \; \Psi ^{(j=0)} \;\; \rightarrow \;\;
\left ( \begin{array}{l}
    -f_{4} \\  0  \\ -f_{2} \\  0
\end{array} \right ) = \; N \;
\left ( \begin{array}{r}
  0 \\ f_{2} \\  0  \\ f_{4}
\end{array}  \right )
$$

\noindent respectively. Evidently, they both  have no solutions,
excluding trivially null ones (and therefore being of~no interest). Moreover,
as  may  be easily seen, in both cases a~function $\Phi (x)$,  defined by
$\hat{N}_{sph.} \; \Psi ^{(j =0)}  \equiv  \Phi (x)$,
lies outside a~fixed totality of states that are only valid as allowed quantum
states of the system under consideration.
At greater values of this $k$, we come  to  analogous  relations:
the~equation $\hat{N}_{sph.} \; \Psi _{j_{min.}} = N \;
\Psi _{j_{min.}}$  leads to
\indent positive $k$:
$$
(-1)^{j+1}   \;
\left ( \begin{array}{l}
        0  \\  f_{3} \; D_{k+1/2}   \\ 0 \\  f_{1} \; D_{k+1/2}
\end{array} \right ) = \; N \;
\left ( \begin{array}{l}
         f_{1} \; D_{k-1/2}  \\   0 \\  f_{3} \; D_{k-1/2} \\  0
\end{array} \right )  \; ;
$$

negative $k$:
$$
(-1)^{j+1} \;
\left ( \begin{array}{l}
   f_{4} \; D_{k-1/2} \\ 0  \\  f_{2} \; D_{k-1/2}  \\ 0
\end{array} \right ) \; = \; N \;
\left ( \begin{array}{l}
  0 \\ f_{2} \; D_{k+1/2}  \\  0  \\ f_{4} \; D_{k+1/2}
\end{array}  \right )
$$

\noindent and the same arguments above may be repeated again.

In turn, as regards the operator $\hat{K}^{k}$ , for the $j_{\min }$  states  we
get
$\hat{K}^{k} \; \Psi _{j_{min.}} =  0$ ;
that is, this state represents the  proper  function  of  the $\hat{K}$
with the null  proper  value.  So,  application  of  this $\hat{K}$
instead of  the $\hat{N}$   has  an  advantage  of  avoiding  the
paradoxical and puzzling situation when $\hat{N}_{sph.} \;
\Psi ^{(j_{min})} \not\in  \{ \Psi  \}$.
In a sense, this second alternative ( the use  of $\hat{K}^{k}$ instead
of $\hat{N}$ at separating the variables and  constructing  the  complete
set of mutually commuting operators) gives us a possibility not to
attach great significance to the monopole discrete operator $\hat{N}$  but
to focus our  attention  solely  on  the  continual  operator $\hat{K}^{k}$.
Indeed, we have described both these alternatives in  case  either
one (first or second) be required.

\subsection*{5. Some additional facts on the monopole system}

Now let  us   consider
relationship between $D$-functions used above  and the
so-called spinor monopole harmonics. To this end one ought to
 perform two translations: from the spherical  tetrad  and
2-spinor (by Weyl) frame in bispinor space  into,  respectively,
the~Cartesian  tetrad and the so-called Pauli's (bispinor) frame.
In the first place, it is convenient to accomplish those translations
for a~free electronic function; so as, in the second place, to follow this
pattern further in the monopole case.

So, subjecting that free electronic  function (spherical solution
from  Sec. 3)   to  the~local  bispinor  gauge   transformation
(associated with  the~tetrad change $e_{sph.} \rightarrow  e_{Cart.}$ )
$$
\Psi _{Cart.} =
\left ( \begin{array}{cc}
     U^{-1}  & 0 \\ 0  & U^{-1}
\end{array} \right )  \; \Psi_{sph.}\; , \qquad U^{-1} =
\left ( \begin{array}{lr}
  \cos \theta /2 \; e^{-i\phi /2}  &  - \sin \theta /2 \; e^{-i\phi /2}  \\
  \sin \theta /2 \; e^{+i\phi /2}  &    \cos \theta /2 \; e^{+i\phi /2}
\end{array} \right )
$$

\noindent and further, taking the bispinor  frame  from  the~Weyl  2-spinor
form into the~Pauli's
$$
\Psi ^{P.}_{Cart.} = \left ( \begin{array}{c}
                 \varphi   \\ \xi
\end{array} \right  ) \; , \qquad
\Psi _{Cart.} = \left ( \begin{array}{c}
             \xi   \\ \eta
\end{array} \right ) \; , \qquad
\varphi  = { \xi  + \eta  \over \sqrt{2}} ,\;\;
\chi     = { \xi -  \eta \over \sqrt{2}}
$$

\noindent we get
$$
\varphi  = \left [ \; {f_{1} + f_{3} \over \sqrt{2} } \;
\left ( \begin{array}{c}
    \cos \theta /2 \; e^{-i\phi /2}  \\
    \sin \theta /2 \; e^{+i\phi /2}
\end{array} \right ) \; D_{-1/2} \; +  \;
{ f_{2} + f_{4} \over \sqrt{2}}
\left ( \begin{array}{c}
   -\sin \theta /2 \; e^{-i\phi /2} \\
    \cos \theta /2 \; e^{+i\phi /2}
\end{array} \right )  \; D_{+1/2} \; \right ]\;  ;
\eqno(A.1a)
$$
$$
\chi  = \left [ \; {f_{1} - f_{3} \over \sqrt{2} } \;
\left ( \begin{array}{c}
    \cos \theta /2 \; e^{-i\phi /2}  \\
    \sin \theta /2 \; e^{+i\phi /2}
\end{array} \right ) \; D_{-1/2} \; +  \;
{ f_{2} - f_{4} \over \sqrt{2}}
\left ( \begin{array}{c}
   -\sin \theta /2 \; e^{-i\phi /2} \\
    \cos \theta /2 \; e^{+i\phi /2}
\end{array} \right )  \; D_{+1/2}\; \right ] \; .
\eqno(A.1b)
$$

\noindent Further, for the~above solutions with fixed proper  values  of
$\hat{\Pi}$-operator, we produce
$$
\Pi = (-1)^{j+1} \; : \;\;\;  \Psi ^{P.}_{Cart.} =
{e^{-i\epsilon t} \over r \sqrt{2}} \;
\left ( \begin{array}{c}
(f_{1} + f_{2})\; (\; \chi _{+1/2} \; D_{-1/2}\; +\; \chi _{-1/2} \;D_{+1/2}\; ) \\
(f_{1} - f_{2})\; (\; \chi _{+1/2} \; D_{-1/2} \;-\; \chi _{-1/2} \; D_{+1/2}\;)
\end{array} \right ) \; ,
\eqno(A.2a)
$$
$$
\Pi =   (-1)^{j}\; : \;\;\; \Psi ^{P.}_{Cart.} =
{ e^{-i\epsilon t} \over  r \sqrt{2}} \;
\left ( \begin{array}{c}
(f_{1} - f_{2})\; (\; \chi _{+1/2} \; D_{-1/2}\; -\; \chi _{-1/2} \; D_{+1/2}\;) \\
(f_{1} + f_{2})\; ( \chi _{+1/2} \; D_{-1/2} \;+\; \chi _{-1/2} \; D_{+1/2}\; )
\end{array} \right )
\eqno(A.2b)
$$

\noindent where $\chi _{+1/2}$   and $\chi _{-1/2}$   designate   the~colomns
of   matrix $U^{-1}(\theta ,\phi )$ (in the literature they are termed as
helicity spinors)
$$
\chi _{+1/2} =
\left ( \begin{array}{c}
    \cos \theta /2 \;  e^{-i\phi /2} \\
    \sin \theta /2 \;  e^{+i\phi /2}
\end{array} \right ) \; , \qquad
\chi _{-1/2} =
\left ( \begin{array}{c}
          -\sin \theta /2  \; e^{-i\phi /2}  \\
           \cos \theta /2  \; e^{+i\phi /2}
\end{array} \right ) \; .
\eqno(A.2c)
$$

\noindent Now, using the  known extensions  for  spherical  spinors
$\Omega ^{j\pm 1/2}_{jm}(\theta ,\phi )$  in terms of
$\chi _{\pm 1/2}$  and  $D$-functions [66]:
$$
\Omega ^{j+1/2}_{jm} = (-1)^{m+1/2} \sqrt{(2j+1)/8\pi}\;
\left [\; \; \chi _{+1/2} \; D_{-1/2} \; + \; \chi _{-1/2} \; D_{+1/2}\;
\right ] \; ,
$$
$$
\Omega ^{j-1/2}_{jm} = (-1)^{m+1/2} \sqrt{(2j+1)/8\pi} \;
\left [\;
- \chi _{+1/2} \; D_{-1/2} \; + \; \chi _{-1/2} \; D_{+1/2} \; \right ]
$$

\noindent we   eventually   arrive   at   the~common   representation   of
the~spinor spherical solutions
$$
\Pi = (-1)^{j+1}\; : \qquad \Psi ^{P.}_{Cart.}=
{e^{-i\epsilon t} \over  r} \;
\left ( \begin{array}{r}
   + f(r) \; \Omega ^{j+1/2}_{jm} (\theta ,\phi ) \\
   - i\;g(r)\; \Omega ^{j-1/2}_{jm}(\theta ,\phi )
\end{array} \right )   \;    ;
\eqno(A.3a)
$$
$$
\Pi = (-1)^{j}\; : \qquad \Psi ^{P.}_{Cart.} =
{e^{-i\epsilon t} \over r} \;
\left ( \begin{array}{r}
-i\; g(r) \; \Omega ^{j-1/2}_{jm} (\theta ,\phi ) \\
     f(r) \; \Omega ^{j+1/2}_{jm}(\theta ,\phi )
\end{array} \right )    \;   .
\eqno(A.3b)
$$

The monopole situation  can  be  considered in the same  way.
As a~result,  we  produce  the~following  representation  of
the~monopole-electron  functions  in  terms  of   `new'   angular
harmonics
$$
N = (-1)^{j+1}\; : \qquad \Psi ^{P.}_{Cart.} = {e^{-i\epsilon t} \over r} \;
\left ( \begin{array}{r}
 + f(r) \; \xi ^{(1)}_{jmk} (\theta ,\phi ) \\
-i \; g(r) \; \xi ^{(2)}_{jmk}(\theta ,\phi )
\end{array} \right )    \;       ;
\eqno(A.4a)
$$
$$
N = (-1)^{j} \; : \qquad \Psi ^{P.}_{Cart.} ={ e^{-i\epsilon t} \over  r} \;
\left ( \begin{array}{r}
-i \; g(r); \xi ^{(1)}_{jmk}(\theta ,\phi ) \\
+f(r) \; \xi ^{(2)}_{jmk}(\theta ,\phi )
\end{array} \right )  \; .
\eqno(A.4b)
$$

\noindent Here, the two column functions $\xi ^{(1)}_{jmk} (\theta ,\phi )$
and   $\xi  ^{(2)}_{jmk} (\theta ,\phi )$ denote  special  combinations
of $\chi _{\pm 1/2}(\theta ,\phi )$ and
$D_{-m,eg/hc\pm 1/2}(\phi ,\theta ,0)$:
$$
\xi ^{(1)}_{jmk} =  [\; \chi _{-1/2}\; D_{k+1/2}\; +  \;
 \chi _{+1/2} \;D_{k-1/2} \; ] \; , \;\;
\xi ^{(2)}_{jmk} = [\; \chi _{-1/2} \; D_{k+1/2} \; - \;
\chi _{+1/2} \; D_{k-1/2}\; ]
\eqno(A.5)
$$

\noindent compare them with  analogous  extensions  for
$\Omega ^{j\pm 1/2}_{jm}(\theta ,\phi )$.  These
2-component and $(\theta ,\phi )$-dependent functions
 $\xi ^{(1)}_{jmk}(\theta ,\phi)$  and $\xi ^{(2)}_{jmk}(\theta ,\phi)$
just  provide  what is called spinor  monopole   harmonics. It
should be useful to write out the detailed explicit form of these
generalized harmonics.  Given  the  known  expressions  for $\chi $-  and
$D$-functions, the formulas (A.5) yield  the following
$$
\xi ^{(1,2)}_{jmk}(\theta ,\phi ) = \left  [ \;
e^{{\rm im}\phi } \;
\left ( \begin{array}{r}
-\sin \theta /2  \; e^{-i\phi /2}  \\
 \cos \theta /2  \; e^{+i\phi /2}
\end{array} \right ) \;
 d^{j}_{-m,k+1/2} (\cos \theta ) \;   \pm  \right.
$$
$$
\left. e^{im\phi } \;
\left ( \begin{array}{c}
\cos \theta /2 \; e^{-i\phi /2} \\
\sin \theta /2 \; e^{+i\phi /2}
\end{array} \right ) \;
d^{j}_{-m,k-1/2} (\cos \theta ) \; \right ]
\eqno(A.6)
$$

\noindent here, the signs  $+ \; (plus)$  and  $- \; (minus)$ refer  to
$\xi ^{(1)}$ and $\xi ^{(2)}$, respectively.

One can equally work  whether in  terms  of  monopole  harmonics
$\xi ^{(1,2)}(\theta ,\phi )$ or directly in terms of $D$-functions,
but  the  latter
alternative has an~advantage over the former because of
the~straightforward  access  to the "unlimited" $D$-function
apparatus; instead  of proving
and  producing   just disguized  old  results.
In any case, one should
establish  existing  correlations and relations (as much as possible)
between at first sight unrelated matters; namely,  the~tetrad formalism,
 special Schr\"odinger basis, Pauli's investigation [64,65],
$D$-function apparatus, and  spinor  (scalar,  vector,  and  so  on)
harmonics. It should be mentioned that to the above list, we
ought to add the so-called formalism  (of  great  popularity)  of
spin-weight harmonics, which  was developed in  the~light  tetrad
frame (also known as the Newman-Penrose formalism).

Above, at translating the electron-monopole functions into the
Cartesian tetrad and Pauli's spin frame, we had overlooked the case
of minimal $j$. Returning to it, on  straightforward  calculation
we find (for $k\; <\; 0$ and $k\; > \; 0$ , respectively)
$$
positive \;\; \kappa : \qquad
\Psi ^{Cart.}_{j_{min.}} \; = \; {e^{-i\epsilon t} \over \sqrt{2} r} \;
\left ( \begin{array}{c}
   ( f_{1} +  f_{3}) \;  \chi _{+1/2} \\
   ( f_{1} -  f_{3}) \;  \chi _{+1/2}
\end{array} \right ) \;
D^{\mid k \mid -1/2} _{-m,k-1/2} (\theta ,\phi ,0) \; ;
\eqno(A.7a)
$$
$$
negative \;\; \kappa :\qquad
\Psi ^{Cart.}_{j_{min.}} =  {e^{-i\epsilon t} \over \sqrt{2} r} \;
\left ( \begin{array}{c}
   ( f_{2} +  f_{4}) \;  \chi _{-1/2} \\
   ( f_{2} -  f_{4}) \;  \chi _{-1/2}
\end{array} \right ) \;
D^{\mid k \mid -1/2} _{-m,k+1/2} (\theta ,\phi ,0)  \; .
\eqno(A.7b)
$$

Now we pass on to another subject   and  take  up  demonstrating
how the major facts  obtained  so  far  are  extended  to
a~curved background geometry (of spherical symmetry).
All above,  the~flat space monopole potential
 $A_{\phi } = g \cos \theta $  preserves
its  simple form at changing the~flat  space  model  into  a~curved one
of spherical symmetry)
$
A_{\phi } = g \cos \theta  \;\; \rightarrow \;\;
 F_{\theta \phi } = - F_{\phi \theta } = - g \sin \theta
$
and the general covariant Maxwell equation in such a curved space
yields
$$
{1\over \sqrt{-g} } {\partial \over \partial x^{\alpha }}
\sqrt{-g} \; F^{\alpha \beta } = 0 \;\; \rightarrow \;\;
{\partial \over \partial \theta }\left [ e^{\nu +\mu } r^{2} \;
\sin \theta  \; {- g \sin \theta  \over  r^{4} \sin ^{2}\theta  } \;\right ]
\equiv  0 \; .
$$

\noindent So,  the  monopole
potential  (for a~curved background geometry) is given  again as
$A_{\phi } = g \cos \theta$. In a~sequence, the~problem of electron in
extenal monopole field (in a~curved  background)  remains,  in a~whole,
unchanged. There are only some new features  brought  about  by
curvature, but they do not affect the~$(\theta ,\phi )$-aspects of
 the~problem. Thus, we arrive at the~ following
$$
\kappa = +1, +3/2, +2,\ldots \; : \qquad
\Psi ^{k>0}_{j_{min.}}(x) = {1 \over r} \;
\left ( \begin{array}{l}
    f_{1}(r,t) \; D_{k-1/2} \\  0  \\  f_{3}(r,t) \; D_{k-1/2} \\  0
\end{array} \right )
\eqno(A.8a)
$$

\noindent from that  it follows
$$
i e^{-\nu /2} \partial _{t}  f_{1} +
i e^{-\mu /2} \partial_{r}  f_{1} - m f_{3} = 0\;  ,\;\;
i  e^{-\nu /2} \partial _{t} f_{3}  -
i  e^{-\mu /2} \partial_{r} f_{3}  -  m f_{1} = 0
\eqno(A.8b)
$$

\noindent and further
$$
f_{3} = {i \over m} \; \left ( \; e^{-\nu /2} \; \partial_{t} \; + \;
 e^{-\mu /2} \; \partial_{r} \; \right ) \; f_{1}(r,t) \; ,
$$
$$
\left [\; (\; e^{-\nu /2}\; \partial_{t}\; - \;
 e^{-\mu /2} \;\partial _{r} \; ) \;\;
  (\; e^{-\nu /2} \;\partial_{t} \; + \; e^{-\mu /2} \;\partial _{r}\;  )
 \;  + \;  m^{2} \; \right ]\;  f_{1} = 0  \; .
\eqno(A.8c)
$$

\noindent   The  case  $\Psi  ^{k<0} _{j_{min.}}(x)$  can   be
considered in  the  same  way. Let us discuss several simple examples.

\centerline{SPHERICAL GEOMETRY}

\noindent In the spherical coordinates
$$
dS^{2} = \left [\; (dt)^{2} - {(dr)^{2} \over 1 -r^{2} }  -
r^{2} ( (d\theta )^{2} + \sin ^{2}\theta  (d\phi )^{2})\;\right ]
$$

\noindent the equation for $f_{1}(t,r)$  takes the form
$$
[\; (\partial_{t} - \sqrt{1- r^{2}} \; \partial_{r} )\;
    ( \partial_{t} + \sqrt{1-r^{2} } \; \partial_{r} )\;  +\;
 m^{2}\;  ]\;  f_{1} = 0 \; .
$$

\noindent Factorizing $f_{1}$  according to
$f_{1} = e^{-i\epsilon t} f(r)$  and introducing
the~variable $\chi $  by relation  $\sin \chi  = r$
(the metric above  becomes
 $dS^{2} = [(dt)^{2} - (d\chi )^{2} -
\sin ^{2} \chi  ( (d\theta )^{2} + \sin ^{2}\theta  (d\phi )^{2}) ]$),
we get
$$
\left [ { d^{2} \over d\chi ^{2}} \;  + \; ( m^{2} -
\epsilon ^{2})\;\right ]\; f(r) = 0 \;  , \qquad
 f = \exp ({\pm \sqrt{m^{2} - \epsilon ^{2}} \;   \chi}) \; .
$$

\noindent Here, the variable $\chi $  lies in  the
 $[0,\; \pi  ]$  or $[ 0 ,\; \pi /2 ]$
intervals  according to whether the spherical  or  elliptic  space
model is  meant. Else one example is
\begin{center}{LOBACHEVSKI GEOMETRY}\end{center}

\noindent  Here, instead  of the above there will be
$$
r  = \sinh  \chi \; , \qquad f_{1} = e^{-i\epsilon t} \; f(r)\; ,
$$
$$
dS^{2} = \left [\; (dt)^{2} - {(dr)^{2} \over 1+r^{2}} -
r^{2} ( (d\theta )^{2} + \sin ^{2}\theta  (d\phi )^{2})\;\right ]\; ;
$$
$$
dS^{2} = [ \; (dt)^{2} - (d\chi )^{2} -
 \sinh ^{2}\chi  ( (d\theta )^{2} + \sin ^{2}\theta  (d\phi )^{2}) \; ]\; ;
$$
$$
\left [\; {d^{2} \over d\chi ^{2}}  +
(m^{2} - \epsilon ^{2}) \; \right] \; f(r) = 0  \; ,
\;\; \rightarrow \;\;  f =
\exp({\pm \; \sqrt{m^{2} - \epsilon ^{2}  \; \chi } })   \; .
$$

Now, we pass on another  interesting  peculiarity  that concern
properties   of the electron current $J_{\alpha }(x)$. This current is given
by
$
J^{\alpha }(x) = \Psi^{+}(x)\;  \gamma ^{0} \;
\gamma ^{\alpha }(x) \; \Psi (x) \; $.
Noting the wave function substitution
$$
\Psi (t,r,\theta ,\phi ) = { e^{-(\nu +\mu )/4} \over r} \;
\left ( \begin{array}{r}
f_{1}(t,r) \; D_{k-1/2}  \\
f_{2}(t,r) \; D_{k+1/2}  \\
f_{3}(t,r) \; D_{k-1/2}  \\
f_{4}(t,r) \; D_{k+1/2}
\end{array} \right )
\eqno(A.9a)
$$

\noindent for  those current components we  get
$$
J^{t}(x) = e^{t}_{(0)} \;\left  [\;
d^{2} _{k-1/2}(\theta ) \; (\mid f_{1}\mid^{2} \; +
\;\mid f_{3}\mid^{2}) \; + \; d^{2} _{k+1/2}(\theta \; ) \;
( \; \mid f_{4}\mid^{2} \; + \;\mid f_{2} \mid^{2} \; )  \; \right ] \; ,
$$
$$
J^{r}(x) = e^{r}_{(3)}\;
\left [\; d^{2}_{k-1/2} (\theta ) \; ( \; \mid f_{1}\mid^{2} \; - \;
\mid f_{3} \mid^{2} \;) \; + \; d^{2}_{k+1/2} ( \theta ) \;
 (\; \mid f_{4}\mid^{2} \; - \;\mid f_{2} \mid^{2} \;) \;\right ] \; ,
$$
$$
J^{\theta }(x) = e^{\theta }_{(1)} \;
\left [ \; (\; f^{*}_{1} \; f_{2}  \; + \;  f_{1} \; f^{*}_{2}\; )\; - \;
(\; f^{*}_{3} f_{4}  \; + \; f_{3}  f^{*}_{4}) \; ) \;
d_{k-1/2}(\theta ) \; d_{k+1/2}(\theta ) \; \right ] \;  ,
$$
$$
J^{\phi }(x) \; =\; -i \;e^{\phi }_{(2}  \;
\left [\; (\; f^{*}_{1} \; f_{2} \; - \; f_{1} \;f^{*}_{2}\; ) \;  -
\; (\; f^{*}_{3} \; f_{4} \; - \; f_{3} \; f^{*}_{4} ) \; \right ] \;
d_{k-1/2}(\theta ) \; d_{k+1/2} (\theta )
\eqno(A.9b)
$$

\noindent here and in the following, the factor
 $r^{-2} \; e^{-(\nu +\mu )/2})$
is omitted; also  we have taken into account the notation
$$
D_{\sigma }  = D^{j}_{-m,\sigma }(\theta ,\phi ,0) =
e^{-i m \phi }\; d^{j}_{-m,\sigma }(\theta ) \; = \;
e^{-i m \phi}\; d_{\sigma }(\theta ) \; .
$$

\noindent Further, for solutions of fixed $N$-parity  values,
the formulas (A.9b)  result in
$$
N = (-1)^{j} , (-1)^{j+1} \;  : \qquad
J^{t}(x) = \; e^{t}_{(0)} \;
\left [ \; d^{2}_{k-1/2} (\theta ) \; + \;
     d^{2}_{k+1/2} ( \theta) \; \right ] \;
( \; \mid f_{1}\mid^{2} \; + \; \mid f_{3}\mid^{2} \; ) \; ,
$$
$$
J^{r}(x) = \; e^{r}_{(0)} \;
\left [\; d^{2}_{k-1/2}(\theta )  \; + \; d^{2}_{k+1/2}\;\right ] \;
(\; \mid f_{1}\mid^{2} \;  - \; \mid f_{3}\mid ^{2} \; )  \;  ,
$$
$$
J^{\theta }(x) = 0 \; , \qquad  J^{\phi }(x) = -2\;i\; e^{\phi }_{(2)} \;
d_{k-1/2}\;  d_{k+1/2} \; (\; f^{*}_{1} \; f_{2} \; - \; f_{1} \; f^{*}_{2}\;)\; .
\eqno(A.9c)
$$

\noindent In turn, for the $j_{\min }$ states we obtain
$$
k = +1/2,+1,+3/2,\ldots \; : \qquad
J^{t}(x) = \; e^{t}_{(0)} \;  d^{2}_{k-1/2}(\theta ) \;
 (\; \mid f_{1}\mid^{2} \; + \; \mid f_{3}\mid^{2}\; ) \;  ,
$$
$$
J^{r}(x) = \; e^{r}_{(3)} \; d^{2}_{k-1/2}(\theta ) \;
(\; \mid f_{1}\mid^{2} \; - \; \mid f_{3}\mid^{2} \; ) \;  ,
\qquad   J^{\theta }(x) = 0 \; , \qquad   J^{\phi }(x) = 0 \; ;
\eqno(A.10a)
$$
$$
k = -1/2,-1,-3/2,\ldots \; : \qquad
J^{t}(x) = e^{t}_{(0)} \;  d^{2}_{k+1/2}(\theta ) \;
(\;\mid f_{4}\mid^{2} \;  + \; \mid f_{2}\mid^{2} \; )\; ,
$$
$$
J^{r}(x) \; = \; e^{r}_{(3)} \; d^{2}_{k+1/2}(\theta )\;
 (\; \mid f_{4}\mid ^{2} \; - \; \mid f_{2}\mid ^{2} \; ) \; ,  \qquad
J^{\theta }(x) = 0 \; , \qquad  J^{\phi }(x) = 0 \; .
\eqno(A.10b)
$$

\noindent It should be noted that  the $J^{\phi }$ vanishes at
 $j = j_{\min }$.
This   sharply  contrasts  with  behaviour  of $J^{\phi }$  component  for all
remaining values $j$ and also contrasts with  free electronic states (in
the~absence an external monopole potential).

Finally, let us consider the question  of  gauge  choice  for
description  of  the  monopole  potential.
From  general considerations we can conclude that, for the problems
considered  above,  it  was  not  basically   essential  thing
whether to use the Schwinger's form of the~monopole  potential  or
to use any other form. Every possible choice  could  bring  about  some
technical incidental variation in a~corresponding description,
but thiswill not affect the applicability  of $D$-function
apparatus to the procedure of separating out the~variables
$\theta ,\phi $  in the electron-monopole sustem. For example,  in
the~Dirac gauge the~monopole potential is given by
$$
(A_{a})^{D.}= \left ( \; 0 \; , \; g\; { \vec{n} \times \vec{r} \over
 r \; ( r + \vec{n} \; \vec{r})} \;\right )
\eqno(A.11a)
$$

\noindent which, after translating to spherical coordinates, becomes
$$
A_{\alpha }^{D.}= (\;  A_{t} = 0, A_{r} = 0 ,
A_{\theta } = 0 , A_{\phi } = g (\cos \theta  - 1)\; ) \; .
\eqno(A.11b)
$$

\noindent On comparing $A^{D.}_{\phi }$ with $A^{S.}_{\phi}$, it follows
immediately  that we can relate these electron-monopole  pictures
($S.$ and  $D.$ gauges)  by gauge transformation
$ S(\phi ) = e^{+ik\phi } \;$:
$$
\Psi ^{D.}(x) \; = \; S(\phi ) \; \Psi ^{S.}\; , \qquad
A^{D.}_{\beta }(x) \; =  \; A^{S.}_{\beta }(x) \; -  \;
i\; (\hbar c/e) \; S(\phi ) \; \partial _{\beta} \; S^{-1}(\phi ) \; .
\eqno(A.11c)
$$

\noindent Simultaneously translating the operators
 $\hat{J}^{k}_{j}, \; \hat{K}, \; \hat{N}$   from  $S.$
to $D.$ gauge
$$
\hat{J}^{D.}_{j} = S \; J^{S.}_{j} \; S^{-1} \; , \qquad
\hat{K}^{D.} = S\; K^{S.} \; S^{-1} , \qquad
\hat{N}^{D.}= S \; \hat{N}^{S.} \; S^{-1}
$$

\noindent we produce
$$
\hat{J}^{D.}_{1} = \left (\; l_{1} + {\cos \phi  \over \sin \theta } \;
 ( i \sigma ^{12} - k (1-\cos \theta ) )\; \right )\; ,
$$
$$
\hat{J}^{D.}_{2} = \left ( \; l_{2} + {\sin \phi  \over \sin \theta }\;
 ( i \sigma ^{12} - k (1 - \cos \theta )) \; \right ) \;, \qquad
 j^{D.}_{3}  = (l_{3} - k )  \; ,
$$
$$
\hat{K}^{D.} = - i \; \gamma ^{0}\; \gamma ^{3}\;
 \left (\; i\gamma ^{1} \; \partial _{\theta } + \gamma ^{2} \;
{ i\partial _{\phi } + k + (i\sigma ^{12} - k )
\cos \theta \over \sin \theta} \; \right )  \; ,
$$
$$
\hat{N}^{D.} = e^{ik(2\phi + \pi )} \; \hat{N}^{S.} \;  .
\eqno(A.11d)
$$

\noindent Thus, the explicit forms of the operators vary from  one  representation
to another, but their proper values remain unchanged;  any alterations
in operators and corresponding modifications  in  wave   functions
cancel out  each other  completely.  That is, as it certainly might expected,
the~complete set of proper values provides such a~description that is
invariant, by its implications, under  any possible $U(1)$ gauge transformations.

Now, let us  consider  else  one  variation  in $U(1)$  gauge,
namely, from  Schwinger's gauge to  the~Wu-Yang's.
In the~Wu-Yang  (hereafter, designated as W-Y)  gauge,
the~monopole potential is characterized by two different respective
expressions  in  two complementary spatial regions
$$
0 \le  \theta  <  (\pi /2 + \epsilon ) \;\; \Longrightarrow \;\;
A^{(N)}_{\phi } = g (\cos \theta  - 1)  \; ,
$$
$$
(\pi /2 - \epsilon ) <  \theta  \le  \pi \;\; \Longrightarrow \;\;
A_{\phi }(S) = g (\cos \theta  + 1) \; ,
\eqno(A.12a)
$$

\noindent and the transition from  the $S.$-basis into $W-Y$'s  can be
obtained  by
$$
\Psi ^{S}(x)\; \Longrightarrow \; \Psi ^{W-Y}(x) =
\left \{  \begin{array}{l}
\Psi ^{(N)}(x) \; = \; S^{(N)}(\phi ) \; \Psi ^{S.}(x)\;\; ,  \qquad
 S^{(N)}(\phi ) \; = \; e^{+ik\phi } \\
\Psi ^{(S)}(x) \; = \; S^{(S)}(\phi ) \; \Psi ^{S.}(x) \;\; ,  \qquad
S^{(S)}(\phi ) \; = \; e^{-ik\phi}
\end{array} \right.
\eqno(A.12b)
$$

\noindent Correspondingly,
for the operators $\hat{J}^{k}_{j}, \; \hat{K},\; \hat{N}$   we     get
two different  forms in $N$- and $S$-regions, respectively:
$$
\hat{J}^{\pm }_{1} = \left ( \; l_{1} + {\cos \phi  \over \sin \theta } \;
 (i\sigma ^{12} - k (1 \pm  \cos \theta ))\;\right ) \; ,
$$
$$
\hat{J}^{\pm }_{2}\  = \left (\; l_{2} + {\sin \phi \over \sin \theta } \;
 ( i\sigma ^{12}  - k (1 \pm  \cos \theta )) \;\right  ) \; , \;\;
 j^{D.}_{3} = (l_{3} \pm  k ) \; ,
$$
$$
\hat{K}^{\pm } = - i \; \gamma ^{0}\; \gamma ^{3}  \;
\left ( \;  i \; \gamma ^{1} \partial _{\theta } + \gamma ^{2}\;
{ i\partial _{\phi } \mp k + (i\sigma ^{12} - k) \cos \theta  \over
\sin \theta }\; \right ) \;  ,
$$
$$
\hat{N}^{\pm } = \exp ( \mp i k (2\phi  +\pi ) )\; \hat{N}^{S.}
\eqno(A.12c)
$$

\noindent where the over sign ($+$ or $-$ ) relates to $S.$-region,
 and  the lower one ($-$ or $+$, respectively) to $N.$-region.

It should be noted  that  only  the  Schwinger's $U(1)$  gauge
(in virtue of the relation $\hat{j}_{3} = - i \partial _{\phi } )$) represents
analogue  of  the~Schr\"odinger's (tetrad) basis discussed in Sec.2,
whereas the~Dirac and Wu-Yang gauges are not. The explicit form
of the~third component of a~total conserved momentum
$
J_{3} = - i \;\partial _{\phi } \equiv  j^{Schr.}_{3}
$
can be regarded as a~determining characteristic,
which  specifies  this basis (and its possible generelazations).
The situations in  $S., \; D.$, and $W-Y$ gauges  are as follows
$$
J^{S.}_{3} = l_{3} \; , \qquad  J^{D.}_{3} = ( l_{3} - k) \; , \qquad
J^{(N)}_{3}= ( l_{3} - k )\; , \qquad J^{(S)}_{3} = ( l_{3} + k )\; ,
$$

\noindent what  proves the above assertion.

\end{document}